\documentclass[conference]{IEEEtran}
\IEEEoverridecommandlockouts
\usepackage{cite}
\usepackage{amsmath,amssymb,amsfonts}
\usepackage{algorithmic}
\usepackage{graphicx}
\usepackage{float}
\usepackage{subcaption}
\usepackage{textcomp}
\usepackage{xcolor}
\usepackage{url}
\usepackage{listings}
\usepackage{stfloats}
\usepackage{comment}
\usepackage{hyperref}
\usepackage{verbatim}
\usepackage{caption}
\usepackage{xspace}
\usepackage{threeparttable}
\usepackage{makecell}
\usepackage{cuted}
\usepackage{multirow}  

\def\BibTeX{{\rm B\kern-.05em{\sc i\kern-.025em b}\kern-.08em
    T\kern-.1667em\lower.7ex\hbox{E}\kern-.125emX}}

\begin{document}

\title{
  A Framework for Ethical Judgment of Smart City Applications
}

\author{
\IEEEauthorblockN{Weichen Shi}
\IEEEauthorblockA{\textit{Department of Computer Science, North Carolina State University} \\
Raleigh, NC \\
wshi6@ncsu.edu}
}

\maketitle

\begin{abstract}
As modern cities increasingly adopt a variety of sensors and Internet of Things (IoT) technologies to collect and analyze data about residents, environments, and public services, they are fostering greater interactions among smart city applications, residents, governments, and businesses. This trend makes it essential for regulators to focus on these interactions to manage smart city practices effectively and prevent unethical outcomes. To facilitate ethical analysis for smart city applications, this paper introduces a judgment framework that examines various scenarios where ethical issues may arise. Employing a multi-agent approach, the framework incorporates diverse social entities and applies logic-based ethical rules to identify potential violations. Through a rights-based analysis, we developed a set of 13 ethical principles and rules to guide ethical practices in smart cities. We utilized two specification languages, Prototype Verification System (PVS) and Alloy, to model our multi-agent system. Our analysis suggests that Alloy may be more efficient for formalizing smart cities and conducting ethical rule checks, particularly with the assistance of a human evaluator. Simulations of a real-world smart city application demonstrate that our ethical judgment framework effectively detects unethical outcomes and can be extended for practical use.
 \end{abstract}


\section{Research Purpose and Questions}
\label{sec:questions}

Ethics defines the rules of morally acceptable social practices and behaviors from a general viewpoint. When considering the comprehensive relationship between humans, society, and technology, ethical guidelines can be distilled to regulate software development and the social application of new technologies. This ensures that technology takes into account the potential risks and impacts arising from its interactions with people and society, thereby protecting the rights of stakeholders to a certain extent.
As the Internet of Things (IoT) technology continues to develop and be deployed in smart cities, it is imperative for technical developers, regulators, and researchers to broaden their perspectives beyond mere technological innovation and deployment. They should incorporate ethical considerations into social computing research and software practices.
This paper aims to develop an ethical judgment and control framework for smart city applications from a multi-agent perspective. For existing applications deployed in smart cities, this framework can assist in determining whether they violate ethical rules or pose ethical risks from multiple viewpoints.

The three main research questions are:

    
\textbf{Q1. Acquisition of Rules}: From where and in what ways can we create or learn ethical rules for smart city applications?

\textbf{Q2. Representation}: How can we effectively formalize all social entities relevant to the smart city application based on ethical rules to ensure their applicability?

\textbf{Q3. Process}:  How will the judgment and control framework function and apply the rules to detect any violations?


\section {Background Introduction}
\label{sec:background}

\subsection{Smart City and IoT Applications}

A smart city represents an emerging framework for the future of urban environments, characterized by an intelligent network of connected objects that transmit data using wireless technologies or cloud computing. Typically realized through Internet of Things (IoT) technology, smart cities leverage and integrate smart sensors, data transmission, storage, and cloud technologies. For a smart city to succeed, its IoT components must ensure pervasive wireless connectivity, trustworthy security, and adaptability and flexibility for deployment and extension. According to a report by NIST,~\cite{nist} the key goals of smart cities include: 1. Faster and broader delivery of urban services; 2. Reduction in the costs of operating a resident-responsive infrastructure; 3. Increased opportunities for interaction, collaboration, and commerce among residents, businesses, and government agencies; 4. Enhanced environmental sustainability; 5. Support for equitable access to city services, including healthcare; 6. Improved quality of life for residents.
Several examples of smart city applications illustrate these goals:

    \textbf{LinkNYC:} A communications network that provides super-fast, free public Wi-Fi, phone calls, device charging, and access to city services via a tablet.
    
    \textbf{Alibaba’s City Brain:} An AI initiative designed to resolve traffic issues in Hangzhou, China, by utilizing data from the transportation bureau, public transport systems, a mapping app, and hundreds of thousands of cameras.
    
    \textbf{Amsterdam Smart City: } A project that opens data vaults, including mapping apps based on shared traffic and transportation data, autonomous delivery boats, and other city management technologies.

Numerous companies are also researching and developing smart city solutions. For example: ~\cite{discoversmartcity}

   \textbf{Athena SmartCities:} A U.S. startup that combines wired and wireless technologies to connect urban elements. Athena SmartESG provides tools for assessing flood risk, while Athena AHOME addresses affordable housing through a digital platform. And the Athena Smart Township Ecosystem Platform is a digital infrastructure that integrates various systems, services, and stakeholders within a community to facilitate sustainable development.
   
    \textbf{AustinGIS: } A U.S.-based startup offering smart infrastructure-as-a-service (IaaS) at scale to city residents, municipalities, and businesses. They also provide public safety services that integrate outdoor cameras, motion sensors, and audible alarms for vehicles and crowds.


\subsection{Limitations on Current Research}

While building a successful smart city necessitates advancements in IoT device development, such as more reliable, secure, and ubiquitous wireless technologies, other crucial aspects must also be considered to achieve the ultimate goals of a smart city. According to the goals outlined by NIST, ~\cite{nist} smart cities should facilitate better interactions among residents, businesses, and governments, as well as ensure equitable access to services and resources. Therefore, there may be limitations in the current discourse surrounding this topic.

A review of the existing research and literature on smart cities and IoT reveals that human and social elements are often overlooked in these analyses. For instance, the U.S. Department of Commerce recommends creating “a stable, secure, and trustworthy IoT environment,”~\cite{ntia} but this focus neglects the social dimensions. When researchers address social aspects of IoT technology and smart cities, they frequently concentrate on topics such as sustainability, behavioral and economic impacts, without emphasizing an ethical perspective. Additionally, the interactions among stakeholders—residents, businesses, and governments, are often ignored.

\subsection{Smart City as a Socio-technical System}

A socio-technical system integrates people and technology, applies an understanding of social structures, roles, and rights to design systems that consider both social and technical factors. In the context of a smart city, framing its social entities and interactions within a socio-technical system can enhance our understanding of the ethical issues that may arise when social entities utilize smart city applications as public services.

This paper on ethical judgment adopts the socio-technical framework for smart cities proposed by M.P. Singh and Murukannaiah. ~\cite{mpsingh2023} Figure 1 illustrates the smart city socio-technical system, which comprises residents, government agencies, and businesses as principles. These entities interact with one another, utilize public services and resources, and produce or consume data.

\begin{figure}[!h]
     \centering
     \includegraphics[width=0.55\textwidth]{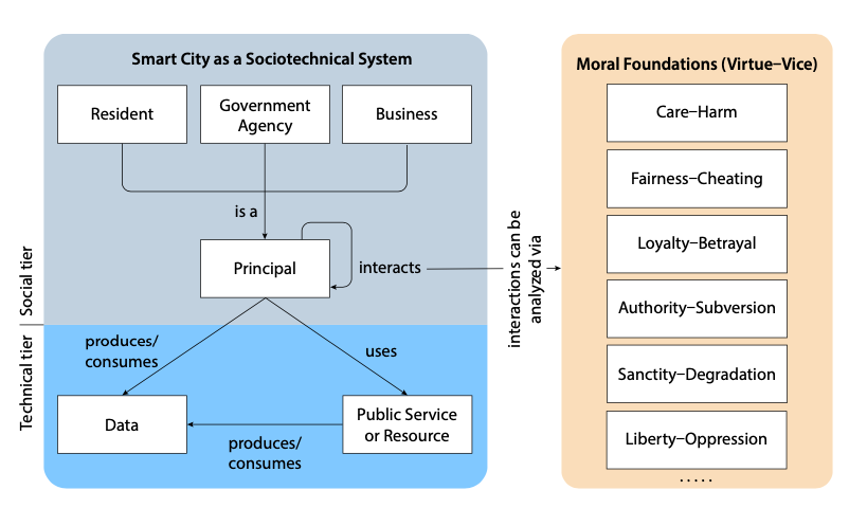}
     \caption{Smart City as a Sociotechnical System, its interactions can be analyzed via moral foundations. M.P. Singh et al, IEEE Internet Computing ’22}
     \label{fig:training_data_shapes}
 \end{figure}
 
\subsection{Ethical Framework and Foundations for Smart City}

To analyze the ethical outcomes that may arise in the interactions of social entities in smart cities, it is essential to propose an ethical framework that encompasses multiple aspects of ethical concerns. This framework enables a comprehensive examination of these interactions.

As illustrated in Figure. 1, the ethical framework adopted in ~\cite{mpsingh2023} applies Moral Foundations Theory (MFT) that includes the following six moral foundations:

\textbf{Care-Harm:} This foundation emphasizes the protection of the vulnerable and reflects people's concern for the well-being of others.

\textbf{Fairness-Cheating:} This principle ensures that individuals are treated fairly.

\textbf{Loyalty-Betrayal:} This foundation pertains to acting in the interest of one’s social group.

\textbf{Authority-Subversion: } This principle involves respecting hierarchies to maintain obedience and deference toward authoritative institutions, such as courts.

\textbf{Sanctity-Degradation: } This foundation relates to feelings of disgust toward contamination.

\textbf{Liberty-Oppression: } This principle focuses on resisting domination and supporting solidarity against oppression.
\\ \hspace*{\fill} \\
However, since the aforementioned Moral Foundations Theory is primarily applicable to analyses in the public policy domain, its weightings may vary in the context of applications in smart cities. For instance, the Sanctity-Degradation foundation may not be relevant when examining interactions among social entities in smart cities. Therefore, this paper adopts a different ethical framework: a rights-based framework. The rights pertinent to the ethical analysis of smart cities include:

\begin{description}
    \item The right to the \textbf{truth} 

    \item The right of \textbf{safety and personal well-being} 

    \item The right of \textbf{privacy} 

    \item The right of \textbf{fairness}

     \item The right of \textbf{what is agreed} 

    \item The right of \textbf{authority} 

\end{description}

\section{Related Work}
\label{sec:related-work}
To build a framework for the ethical judgment of smart cities applications, it is essential to understand how ethical principles can be derived from use cases and formalized as rules, as well as how to apply these rules within a smart city multi-agent system to facilitate ethical judgment. The related work for this project includes smart city IoT applications, ethical studies, ethics formalization, and ethical judgment.

\subsection{Smart City Ethics}
Our survey of the current literature on the ethics of smart cities reveals a number of relevant papers on this topic. Kitchin ~\cite{robk} highlights that data-driven urbanism and commercial products raise significant ethical issues related to privacy, datafication, dataveillance, geosurveillance, and uses of data such as social sorting and anticipatory governance. Ahmad et al.~\cite{ahmad} argue that the AI code of ethics is a critical aspect of smart city applications, appealing to managers in the tech industry. They provide a taxonomy of key ethical issues discussed in academic literature. Victor Chang~\cite{vchang} proposes an ethical framework for big data and smart cities, focusing on ethical and non-ethical issues in big data analytics applications within public transportation systems. Clever, Crago et al. ~\cite{sawyerc} address general ethical aspects and the responsible use of data in smart cities, compiling a list of challenges facing smart city applications. A review of these publications indicates that the majority of the publications center on big data and AI in smart cities, and often neglect a multi-perspective analysis of smart city applications.

Additionally, in the context of ethics specifically related to IoT applications, Allhoff et al.~\cite{Allhoff} survey foundational ethical issues for the Internet of Things, discussing topics such as informed consent, privacy, information security, physical safety, and trust. They emphasize that these ethical issues do not exist in isolation but converge and intersect with one another.

\subsection{Ethics Formalization and Judgment}
Current state-of-the-art research in formally studying ethics aims to understand human moral decision-making to inform the engineering of ethical AI systems and judgment principles. Awad, Levine, Anderson, et al.~\cite{EdmondAwad} propose a framework known as computational ethics, which specifies how the ethical challenges of AI can be partially addressed by incorporating insights from human moral decision-making. The driving force behind this framework is a computational version of reflective equilibrium (RE), an approach that seeks coherence between considered judgments and governing principles. This work offers a novel computational perspective for analyzing ethics within the broader context of AI.

Jacovi et al.~\cite{Jacovi} focus on trust in artificial intelligence, providing several working definitions and formalizations for trust-related terms. They address key questions about the prerequisites for establishing trust, the goals of trust, and what is necessary to achieve those goals. While their work thoroughly explores a specific ethical concept, it does not employ a logic-based formalization method.

For logic-based ethical system formalization, Jean-Gabriel Ganascia~\cite{Ganascia} presents a detailed example of modeling an ethical system using non-monotonic logic. This model utilizes Answer Set Programming (ASP) formalism to extend classical philosophy and define the general conditions required for any moral system.

Cointe et al.~\cite{Cointe} propose a generic judgment capability for autonomous agents, which involves the explicit representation of elements such as moral values, moral rules, and ethical principles. They demonstrate how this model facilitates the comparison of the ethics of different agents. Additionally, this moral judgment model is designed as a modular component that can be integrated into existing architectures to add an ethical layer to the decision-making process. However, the judgment model presented in their work cannot be directly mapped to specific ethical concerns within the ethical framework and may not be applicable to IoT software and interactions in smart cities.
\\ \hspace*{\fill} \\

\section{Overall Ethical Judgment Strategy}
\label{sec:overall-strategy}

The three research questions proposed in~\ref{sec:questions} pertain to the acquisition of ethical rules, the formal representation of social entities, and the processes for ethical judgment and control. Given that specific ethical rules for IoT applications in smart cities involve multiple social entities, such as residents and governments, it is important to study ethics from multiple perspectives within the socio-technical framework of smart cities, as illustrated in Figure 1. Once a set of ethical principles is proposed, we can formalize them into logic rules and apply them for automated ethical judgment. A representation of social entities (agents) and their properties should be formalized before ethical rules could be written.

This project adopts two specification languages, PVS and Alloy, for the formal representation of social entities—namely, IoT services, residents, governments, businesses, and their properties. ~\ref{sec:formalization} provides a detailed introduction to the formalizations of each entity.

For our first research task i.e., the acquisition of ethical rules, this research has considered three different approaches: ~\cite{Cointe}

\textbf{Ethical rules by design:} This approach involves designing ethical rules through a priori analysis of every possible situation where an ethical foundation might be violated or an unethical outcome could occur.

\textbf{Logic-based ethics: } This approach models specific ethical foundations using formal logic, accounting for the various aspects that constitute a given ethical framework.

\textbf{Ethical rules Inference: } This method infers ethical rules from a large set of ethical judgment examples produced by humans.

Due to the often overlapping nature of ethical issues and the challenges associated with directly modeling specific ethical principles using formal logic, this paper primarily focuses on the ethical rules by design approach. Once a comprehensive set of ethical principles is developed, it will be formalized as logical rules and applied within a multi-agent ethical judgment system to perform rules verification for targeted smart city IoT applications. The next section presents the process and results of designing ethical principles based on a rights-based analysis of a smart city socio-technical system.
\\ \hspace*{\fill} \\

\section{Rights-based Analysis for Ethical Principles Design}
\label{sec:rights-based}
The ethical and moral framework for the smart city socio-technical system is based on a rights-based framework. As introduced in ~\ref{sec:background}, this framework includes six foundational rights: the right to the truth, the right to safety and personal well-being, the right to privacy, the right to fairness, the right to what is agreed, and the right of authority. To derive a set of ethical rules grounded in these six rights, we will adopt the ethical rules by design approach to conduct a priori analysis of potential scenarios where these ethical foundations may be violated. The following sections present an analysis aimed at developing various ethical principles. In ~\ref{sec:ethical-rules}, these ethical principles will be formalized as logical rules.
\\ \hspace*{\fill}
\begin{itemize}
\item
The right of \textbf{safety and personal well-being}: 
\\ \hspace*{\fill} \\
Safety and personal well-being are among the most crucial properties and ethical dimensions of smart city applications. In various scenarios, a smart city application may interact with residents either physically or non-physically, potentially posing safety threats, particularly when devices can move or change their physical form. An ethical principle reflecting the right to safety and personal well-being could be stated as follows:
\\ \hspace*{\fill} \\
\textbf{P1:} \textit{The service should not be based on a device that involves high-risk physical movement, change, or transmission of harmful substances or information.}
\\ \hspace*{\fill} \\
When considering interactions between government entities and residents, we can establish an ethical principle indicating that the government should avoid introducing smart city devices into residents' living environments where safety threats may exist. This leads to the following principle:
\\ \hspace*{\fill} \\
\textbf{P2:} \textit{Government should not intend to apply an IoT device to transform the physical living environment of residents while residents may interact directly with the device under safety threats.}
\\ \hspace*{\fill} \\
From the resident-resident perspective, another important ethical principle can be derived. For instance, consider a device used in public transportation that collects data about the current occupancy of a bus or train, making this information available to all riders. In such cases, unethical behaviors, like pickpocketing, could emerge, particularly in crowded settings. Thus, we propose the following principle for safety and personal well-being: 
\\ \hspace*{\fill} \\
\textbf{P3:} \textit{Residents should not get information about the physical locations of other residents or have increased opportunity of physical interactions with other residents by using the IoT device.}
\\ \hspace*{\fill} \\
\item
The right of \textbf{privacy}: 
\\ \hspace*{\fill} \\
Privacy is a fundamental right that enables smart city residents to keep their personal information confidential and to control how it is collected and used. From the perspectives of Residents-Government, Residents-Residents, and Residents-Business, we propose three ethical rules regarding the right to privacy:
\\ \hspace*{\fill} \\
\textbf{P4:} \textit{The IoT device should not collect and/or analyze user data and make that information available to government agencies without consent.}
\\ \hspace*{\fill} \\
\textbf{P5:} \textit{The IoT device should not collect and/or analyze user data and make that information public or available to other residents without consent.}
\\ \hspace*{\fill} \\
\textbf{P6:} \textit{The IoT device should not collect and/or analyze user data and make that information used by businesses or obtained by hackers without consent.}
\\ \hspace*{\fill}
\item
The right of \textbf{fairness}: 
\\ \hspace*{\fill} \\
The right to fairness dictates that residents of a smart city should be treated equitably and benefit indiscriminately from IoT devices. By analyzing various scenarios, we can develop principles that reflect this right from multiple perspectives.

For instance, consider a device deployed on a ferry that records the class of onboard passengers, the ferry's current location, and estimates the arrival time. This information could be transmitted to nearby businesses, such as restaurants, allowing them to tailor their services and menu offerings for first-class passengers before they arrive. This practice could be unfair to regular-class passengers, leading to the following ethical principle:
\\ \hspace*{\fill} \\
\textbf{P7:} \textit{Business such as restaurants should not obtain information such as the arrival time and possible class of customers via the smart city technology.}
\\ \hspace*{\fill} \\
A more common violation of the right to fairness may arise when residents have competing preferences regarding a smart city device. For example, a facial recognition device installed at the entrance of a community apartment complex might enhance security and facilitate resident identification for some, while others may be concerned about their images being used for unintended purposes and oppose its implementation. An ethical principle for this scenario could be:
\\ \hspace*{\fill} \\
\textbf{P8:} \textit{The IoT device should not bring up competing preferences towards its usage among residents, so that the device will not be beneficial to certain residents while causing disadvantages to others.}
\\ \hspace*{\fill} \\
Additionally, the locations where smart city devices are deployed may lead to perceptions of unfairness. An ethical principle addressing this concern could be:
\\ \hspace*{\fill} \\
\textbf{P9:} \textit{The smart city technology should not only be deployed in a small subset of neighborhoods or privileged neighborhoods to facilitate nearby businesses.}
\\ \hspace*{\fill} \\
\item
The right to \textbf{the truth}: 
\\ \hspace*{\fill} \\
The right to the truth encompasses the understanding of the full context surrounding circumstances, purposes, participants, and other vital information regarding an event. In a smart city setting, residents should be entitled to this right. Accordingly, a relevant ethical principle could be:
\\ \hspace*{\fill} \\
\textbf{P10:} \textit{Residents should be informed about the goals and purposes of the smart city project as well as matters that may affect their rights.}
\\ \hspace*{\fill} \\
\item
The right to \textbf{what is agreed}: 
\\ \hspace*{\fill} \\
The right to what is agreed grants residents the entitlement to what has been promised and agreed upon regarding smart city devices. A corresponding ethical principle for this right could be:
\\ \hspace*{\fill} \\
\textbf{P11:} \textit{The IoT device should not collect more data or perform actions beyond what has been consented by the residents.}
\\ \hspace*{\fill} \\
\item
The right of \textbf{authority}: 
\\ \hspace*{\fill} \\
The concept of authority empowers individuals or organizations to control and make decisions on specific matters. In the context of smart city technologies, residents, as users and beneficiaries, should possess certain authority to protect their interests, while governments, as regulators, also hold authority. Accordingly, we propose the following two ethical principles:
\\ \hspace*{\fill} \\
\textbf{P12:} \textit{Residents should have the legitimate authority to request actions to avoid undesirable or unlawful outcomes caused by the IoT service.}
\\ \hspace*{\fill} \\
\textbf{P13:} \textit{Government should oversight or enforce minimum safety standards for the IoT device, or increase security in public to protect the safety, privacy, and well-being of residents.}
\\ \hspace*{\fill} \\
\end{itemize}

\section{Formalization of Agents in Smart Cities}
\label{sec:formalization}

Once we have established a set of ethical principles, the next step is to formalize them into logical rules. This formalization relies on formal models of social entities (agents), including IoT Services, Residents, Government, and Businesses. We utilize two specification languages, PVS and Alloy, for this process. This section begins with an introduction to both languages, followed by formal specifications in PVS and specification constructs in Alloy for each of the four agents.

\subsection{PVS Language}

The Prototyping System (PVS) is a specification language that integrates supporting tools and an automated theorem prover,~\cite{PVSprover} developed by SRI's International Computer Science Laboratory in Menlo Park, California.~\cite{PVSwiki}

PVS is based on a kernel that incorporates Church's type theory along with dependency type extensions, fundamentally representing classic typed higher-order logic. Basic types include user-defined types as well as built-in types such as Booleans, integers, real numbers, and ordinal numbers. Type constructors encompass functions, collections, tuples, records, enumerations, and abstract data types. Constraints can be introduced through predicate subtypes and dependency types; these constrained types may create an obligation for proof during type checking, known as a type-correctness condition (TCC). PVS specifications are organized into parametric theories.

\subsection{Alloy}

Alloy is an open-source declarative specification language and analyzer designed for modeling software systems and verifying whether specific properties hold true. It is particularly useful for expressing complex structural constraints and behaviors within a software system, utilizing first-order logic. Alloy specifications can be validated using the Alloy Analyzer.~\cite{Alloywiki} To formalize the four types of social entities in a smart city, an abstract signature can be declared for each entity using Alloy.

\subsection{Formalization of IoT Services Agents}

An IoT service agent represents specific IoT devices deployed within a smart city context. It should include properties that indicate the device's title, the neighborhoods where it is deployed, its exact location, the type of movement, the interaction type, and the associated risk type. Additionally, the agent should include indicators to determine whether the usage agreement of the IoT device may be violated.
\\ \hspace*{\fill} \\
\textbf{\underline{Alloy Specification:}}
\\ \hspace*{\fill} \\
abstract sig \textbf{IoTService} \{

device\_title : String,		\textcolor{gray}{// Title of the device}
\\ \hspace*{\fill} \\
\textcolor{gray}{// Set of neighborhoods where the device is deployed}

deploy\_neighborhood : set String,	
\\ \hspace*{\fill} \\
\textcolor{gray}{// Exact location of the device, represented as integers}

location : set Int,
\\ \hspace*{\fill} \\
\textcolor{gray}{// Type of movement (e.g., stationary, mobile)}

movement\_type: String,
\\ \hspace*{\fill} \\
\textcolor{gray}{// Type of interaction (e.g., physical, non-physical)}

interaction\_type:  String,
\\ \hspace*{\fill} \\	
\textcolor{gray}{// Associated risk type (e.g., low, medium, high)}

risk\_type: String,
\\ \hspace*{\fill} \\	
\textcolor{gray}{// Indicator of whether the usage agreement is violated}

IoT\_usage\_agreement\_violated: String,
\\ \hspace*{\fill} \\		
\textcolor{gray}{// Set of input ports for communication}

in\_ports: set String,
\\ \hspace*{\fill} \\	
\textcolor{gray}{// Set of output ports for communication}

out\_ports: set String
		
\}
\\ \hspace*{\fill} \\
\textbf{\underline{PVS Specification:}}
\\ \hspace*{\fill} \\

\textbf{IoTService} [DEVICE: TYPE+]: THEORY

BEGIN 

device\_title: string

deploy\_neighborhood: set of [string]

location: TYPE+ = [\# x, y: real \#]

movement\_type : TYPE+ = {stationary, mobile}

interaction\_type: TYPE+ = {physical, non-physical }

risk\_type: TYPE+ = {low, medium, high}

IoT\_usage\_agreement\_violated: bool

IoT\_devices: set of[DEVICE]

in\_ports, out\_ports:set of[PORT]

collect\_residents\_data:  bool

collect\_user\_data: [PERSONALINFO \textrightarrow PORT]

analyze\_user\_data: 

[PERSONALINFO \textrightarrow PERSONALINFO]

send\_user\_info: [[DEVICE,[PORT \textrightarrow PERSONALINFO]] 

\textrightarrow [SOCIALENTITY]]

AgentInfo: TYPE=[\#inter\_state: DEVICE, 

checkpoint:CHECKSTATE, rec:SIG \#]

send\_state\_to\_judge: [[DEVICE,

[PORT \textrightarrow AgentInfo]] \textrightarrow [JUDGE]]

info: VAR AgentInfo

SendStateToCommunicationAgent(info):

Send=IF member(inter\_state(ts), n\_states) 

THEN(\#inter\_state:=JUDGE)\_1(senddata (inter\_state(info),

(LAMBDA p:dc\_msg(p)))), checkpoint:=inter\_state(info), rec:=0 \#) 

END IoTService

\subsection{Formalization of Residents Agents}

A resident agent represents a group of individuals who may use or interact with IoT devices in a smart city context. This agent should include properties that indicate the neighborhoods where residents live, any preferred neighborhoods, their economic status and professions, their preferences for IoT device usage, and whether they possess legitimate authority to request actions aimed at preventing unlawful outcomes from IoT applications.
\\ \hspace*{\fill} \\
\textbf{\underline{Alloy Specification:}}
\\ \hspace*{\fill} \\
abstract sig \textbf{Residents} \{
\\ \hspace*{\fill} \\
\textcolor{gray}{// Set of neighborhoods where residents live}

all\_living\_neighborhoods : set String,		
\\ \hspace*{\fill} \\
\textcolor{gray}{// Set of neighborhoods preferred by residents}

favored\_neighborhoods : set String,	
\\ \hspace*{\fill} \\
\textcolor{gray}{// Set indicating residents' economic status (e.g., low, middle, high)}

economic\_status: set String,
\\ \hspace*{\fill} \\
\textcolor{gray}{// Set of professions held by residents}

professions: set String,
\\ \hspace*{\fill} \\
\textcolor{gray}{// Set indicating residents' preferences for using IoT devices}

preference\_for\_IoT\_usage: set String,
\\ \hspace*{\fill} \\	
\textcolor{gray}{//  Indicator of whether residents have authority to request actions}

has\_legitimate\_authority: String,
\\ \hspace*{\fill} \\	
\textcolor{gray}{// Indicator of whether the usage agreement is violated}

IoT\_usage\_agreement\_violated: String,
\\ \hspace*{\fill} \\		

in\_ports: set String,

out\_ports: set String
		
\}
\\ \hspace*{\fill} \\
\\ \hspace*{\fill} \\
\textbf{\underline{PVS Specification:}}
\\ \hspace*{\fill} \\
\textbf{Residents} [SOCIALENTITY: TYPE+]: THEORY

BEGIN 

Residents : TYPE FROM SOCIALENTITY

all\_living\_neighborhoods: set of [string]

preferred\_neighborhoods: set of [string]

economic\_status: TYPE+ = \{high, middle, low\}

professions: set of [string]

preference\_for\_IoT\_usage: set of [string]

in\_ports, out\_ports:set of[PORT]

has\_legitimate\_authority: bool

collect\_personal\_data: [PERSONALINFO \textrightarrow PORT]

send\_personal\_info\_to\_socialentity: 

[[SOCIALENTITY,[PORT \textrightarrow PERSONALINFO]] 

\textrightarrow [SOCIALENTITY]]

receive\_msg: [[SOCIALENTITY,[PORT 

\textrightarrow PERSONALINFO]] \textrightarrow [SOCIALENTITY]]

AgentInfo: TYPE=[\#inter\_state: SOCIALENTITY, 

checkpoint:CHECKSTATE, rec:SIG \#]

send\_state\_to\_judge: [[SOCIALENTITY,

[PORT \textrightarrow AgentInfo]] \textrightarrow [JUDGE]]

info: VAR AgentInfo

SendStateToCommunicationAgent(info):

Send=IF member(inter\_state(ts), n\_states) 

THEN(\#inter\_state:=JUDGE)\_1(senddata (inter\_state(info),

(LAMBDA p:dc\_msg(p)))), checkpoint:=inter\_state(info), rec:=0 \#) 

END Residents

\subsection{Formalization of Government Agents}

A government agent represents a government agency that may initiate a smart city project and oversee the deployment of IoT applications within that context. This agent should include properties that indicate the type of government entity, the goals of the smart city project, and indicators regarding whether the agency monitors safety and enforces safety standards for the IoT application.
\\ \hspace*{\fill} \\
\textbf{\underline{Alloy Specification:}}
\\ \hspace*{\fill} \\
abstract sig \textbf{Government } \{
\\ \hspace*{\fill} \\
\textcolor{gray}{// Set of types of government entities (e.g., local, state, federal)}

gov\_types : String,		
\\ \hspace*{\fill} \\
\textcolor{gray}{// Set of goals for the smart city IoT project (e.g., sustainability, efficiency)}

IoT\_project\_goals : set String,	
\\ \hspace*{\fill} \\
\textcolor{gray}{// Indicator of whether the agency monitors IoT safety}

oversight\_IoT\_safety:  String,
\\ \hspace*{\fill} \\
\textcolor{gray}{// Indicator of whether the agency enforces safety standards for IoT applications}

enforce\_safety\_standards:  String,
\\ \hspace*{\fill} \\		

in\_ports: set String,

out\_ports: set String
\}
\\ \hspace*{\fill} \\
\textbf{\underline{PVS Specification:}}
\\ \hspace*{\fill} \\
\textbf{Government} [SOCIALENTITY: TYPE+]: THEORY

BEGIN 

Government : TYPE FROM SOCIALENTITY

gov\_types: set of [string]

IoT\_project\_goals: set of [string]

oversight\_IoT\_safety: bool

enforce\_safety\_standards: bool

in\_ports, out\_ports:set of[PORT]

send\_meg: [[SOCIALENTITY,

[PORT \textrightarrow MSG]] \textrightarrow [SOCIALENTITY]]

receive\_msg: [[SOCIALENTITY,[PORT 

\textrightarrow MSG]] \textrightarrow [SOCIALENTITY]]

AgentInfo: TYPE=[\#inter\_state: SOCIALENTITY, 

checkpoint:CHECKSTATE, rec:SIG \#]

send\_state\_to\_judge: [[SOCIALENTITY,

[PORT \textrightarrow AgentInfo]] \textrightarrow [JUDGE]]

info: VAR AgentInfo

SendStateToCommunicationAgent(info):

Send=IF member(inter\_state(ts), n\_states) 

THEN(\#inter\_state:=JUDGE)\_1(senddata (inter\_state(info),

(LAMBDA p:dc\_msg(p)))), checkpoint:=inter\_state(info), rec:=0 \#) 

END Government

\subsection{Formalization of Business Agents}

A business agent represents a business owner who may interact with or leverage IoT applications to enhance their operations in a smart city context. This agent should include properties that specify the scale of the business, the type of business, and the neighborhoods where the business is situated.
\\ \hspace*{\fill} \\
\textbf{\underline{Alloy Specification:}}
\\ \hspace*{\fill} \\
abstract sig \textbf{BA } \{
\\ \hspace*{\fill} \\
\textcolor{gray}{// Set indicating the scale of the business (e.g., "small," "medium," "large")}

scale : set String,		
\\ \hspace*{\fill} \\
\textcolor{gray}{// SSet of neighborhoods where the business operates}

neighborhoods : set String,	
\\ \hspace*{\fill} \\
\textcolor{gray}{// Set specifying the type of business (e.g., "retail," "service," "manufacturing")}

type\_of\_business: set String,
\\ \hspace*{\fill} \\		

in\_ports: set String,

out\_ports: set String
\}
\\ \hspace*{\fill} \\
\textbf{\underline{PVS Specification:}}
\\ \hspace*{\fill} \\
\textbf{BA} [SOCIALENTITY: TYPE+]: THEORY

BEGIN 

BA : TYPE FROM SOCIALENTITY

scale: TYPE+ = \{large, med, small\}

neighborhoods: set of [string]

type\_of\_business: set of [string]

in\_ports, out\_ports:set of[PORT]

send\_meg: [[SOCIALENTITY,

[PORT \textrightarrow MSG]] \textrightarrow [SOCIALENTITY]]

receive\_msg: [[SOCIALENTITY,[PORT 

\textrightarrow MSG]] \textrightarrow [SOCIALENTITY]]

AgentInfo: TYPE=[\#inter\_state: SOCIALENTITY, 

checkpoint:CHECKSTATE, rec:SIG \#]

send\_state\_to\_judge: [[SOCIALENTITY,

[PORT \textrightarrow AgentInfo]] \textrightarrow [JUDGE]]

info: VAR AgentInfo

SendStateToCommunicationAgent(info):

Send=IF member(inter\_state(ts), n\_states) 

THEN(\#inter\_state:=JUDGE)\_1(senddata (inter\_state(info),

(LAMBDA p:dc\_msg(p)))), checkpoint:=inter\_state(info), rec:=0 \#) 

END BA
\\ \hspace*{\fill} \\

\section{Ethical Rules Formalization}
\label{sec:ethical-rules}
This section presents the formalization of ethical principles into logical rules for assessing the ethical outcomes of IoT services in a smart city. In the next page, a pre-designed table outlines the compliance of rights-based ethical principles from multiple perspectives. For each principle, the corresponding formalized logical rules are provided on the right. In ~\ref{sec:case-study}, a real-world case study will offer a detailed analysis of nuanced ethical rules and demonstrate how the ethical rules outlined in this section can be applied in practice.

\begin{table*} [h!]
      \centering
      \begin{tabular}{|c|>{\centering\arraybackslash}p{2cm}|>{\centering\arraybackslash}p{5cm}|>{\centering\arraybackslash}p{8cm}|} \hline 
           Ethical Right&  Perspective&  Principles& Formal Rules\\ \hline 
         \multirow{3}{*}{Safety} &  Residents – IoT Service&  P1 - The service should not be based on a device that involves high-risk physical movement, change, or transmission of harmful substances or information& M(d): The device involves physical movement. \newline H(d): The associated risk level is high. \newline T(d): The device transmits harmful substances or information.\newline $\forall d (Device(d) \implies \neg(M(d) \lor H(d) \lor T(d)))$ \\ \cline{2-4} 
           &  Residents - Government&  P2 - Government should not intend to apply an IoT device to transform the physical living environment of residents while residents may interact directly with the device under safety threats& $\neg("transform\_physical\_living\_environment"
\in IoT\_project\_goals(g,d)) \lor interaction\_type(d)="non\_physical"$
\\ \cline{2-4} 
           &  Residents - Residents&  P3 - Residents should not get information about the physical locations of other residents or have increased opportunity of physical interactions with other residents by using the IoT device& $\neg \exists r \in Residents: \exists l\in Locations : receive\_msg(r,[PORT$\textrightarrow$l]) $ \\ \hline 
           \multirow{3}{*}{Privacy} &  Residents - Government&  P4 - The IoT device should not collect and/or analyze user data and make that information available to government agencies without consent&$
            \neg collect\_residents\_data \lor 
\neg receive\_msg:[[Residents,[PORT$ \textrightarrow $ MSG]]
$\textrightarrow$[Government]]$ \\ \cline{2-4} 
           &  Residents - Residents&  P5 - The IoT device should not collect and/or analyze user data and make that information public or available to other residents without consent& $
            \neg collect\_residents\_data \lor 
\neg receive\_msg:[[Residents,[PORT$ \textrightarrow $ MSG]]
$\textrightarrow$[Residents]]$\\ \cline{2-4} 
           &  Residents - Business&  P6 - The IoT device should not collect and/or analyze user data and make that information used by businesses or obtained by hackers without consent& $
            \neg collect\_residents\_data \lor 
\neg receive\_msg:[[Residents,[PORT$ \textrightarrow $ MSG]]
$\textrightarrow$[BA]]$\\ \hline 
           \multirow{3}{*}{Faireness} &  Residents- Business&  P7 - Business such as restaurants should not obtain information such as the arrival time and possible class of customers via the smart city technology & $
\neg receive\_msg:[[Residents,[PORT$ \textrightarrow $ ARRIVALTIME]]
$\textrightarrow$[Business]] \land \neg receive\_msg:[[Residents,[PORT$ \textrightarrow $ economic\_status]]
$\textrightarrow$[Business]]$ \\ \cline{2-4} 
           &  Residents - Residents&  P8 - The IoT device should not bring up competing preferences towards its usage among residents, so that the device will not be beneficial to certain residents while causing disadvantages to others& size (preference\_for\_IoT\_usage) =1\\ \cline{2-4}
           &  Business - Business&  P9 - The smart city technology should not only be deployed in a small subset of neighborhoods or privileged neighborhoods to facilitate nearby businesses & size(deploy\_neighborhood)  $\geq$ \#total\_num\_neighborhood*0.4 \\ \hline 
 The Truth & Residents - Government& P10 - Residents should be informed about the goals and purposes of the smart city project as well as matters that may affect their rights&$
receive\_msg:[[Government,[PORT$ \textrightarrow $ IoT\_project\_goals]]
$\textrightarrow$[Residents]]=True$\\ \hline 
 What's Agreed & Residents – IoT Service& P11 - The IoT device should not collect more data or perform actions beyond what has been consented by the residents&IoT\_usage\_agreement\_violated = FALSE\\ \hline 
 \multirow{2}{*}{Authority} & Residents – IoT Service& P12 - Residents should have the legitimate authority to request actions to avoid undesirable or unlawful outcomes caused by the IoT service&has\_right\_for\_protection = TRUE\\ \cline{2-4}
 & Residents - Government& P13 - Government should oversight or enforce minimum safety standards for the IoT device, or increase security in public to protect the safety, privacy, and well-being of residents &oversight\_IoT\_satefy = TRUE $\land$ enforce\_safety\_standard = TRUE\\ \hline
      \end{tabular}
      \caption{Ethical principles and formalized rules}
      \label{tab:my_label}
  \end{table*}

The defined rules can be consolidated and represented in PVS specification using a RULESET Theory, for example: 
RULESET EthicalPrinciples = 

  [SafetyPrinciple: (IoTDevice) \textrightarrow BOOL, 
  
   PrivacyPrinciple: (Resident, IoTDevice) \textrightarrow BOOL, 
   
   FairnessPrinciple: (Resident, Business) \textrightarrow BOOL] 
\\ \hspace*{\fill} \\
  SafetyPrinciple(d) == 
  
    (d.movement\_type != hazardous) $\implies$ 
    
    (d.interaction\_type != physical)
\\ \hspace*{\fill} \\
  PrivacyPrinciple(r, d) == 
  
    (d.data\_collection) $\implies$  
    
    (r.informed\_consent)
\\ \hspace*{\fill} \\
  FairnessPrinciple(r, b) == 
  
    (b.service\_access) $\implies$  
    
(r.equal\_opportunity)

In this PVS specification, we define a RULESET called EthicalPrinciples that includes three ethical principles: SafetyPrinciple, PrivacyPrinciple, and FairnessPrinciple. Each principle is defined as a function that returns a Boolean value, indicating whether the rule holds for given inputs, such as an IoT device or a resident. This structured approach allows for clear expression and verification of ethical compliance in the smart city context.

In Alloy, the defined rules can be expressed and consolidated in an assertion. Assertions are constraints that are intended to hold true for the model and are checked to ensure consistency, for example:
\\ \hspace*{\fill} \\
assert SafetyPrinciple \{

    all device: IoTDevice $\|$ 
    
        device.movementType != "hazardous" $\implies$ 
        
        device.interactionType != "physical"
\}
\\ \hspace*{\fill} \\
This assertion states that if an IoT device's movement type is not hazardous, then its interaction type should not be physical, ensuring compliance with safety principles. Such assertions help maintain the integrity of the model and verify that the ethical rules are consistently applied.
\\ \hspace*{\fill} \\

\section{Framework for Ethical Judgment}
\label{sec:framework}
This section outlines a comprehensive framework for the ethical judgment of smart city applications, leveraging formal specifications of social entities within a multi-agent system. The architecture for implementing this judgment framework is detailed in Section A. Subsequent sections will explain how various agents interact within a specific smart city application to generate instances of formal specifications and apply ethical rules to identify any violations.

\subsection{Architecture of Ethical Judgment MAS}

The multi-agent architecture consists of distinct agents representing the key social entities: IoT services, residents, government, and businesses. Each agent operates autonomously but collaborates with others to evaluate ethical compliance. The interactions among these agents facilitate the assessment of ethical principles and the detection of potential violations.

\begin{figure}[!h]
     \centering
     \includegraphics[width=0.56\textwidth]{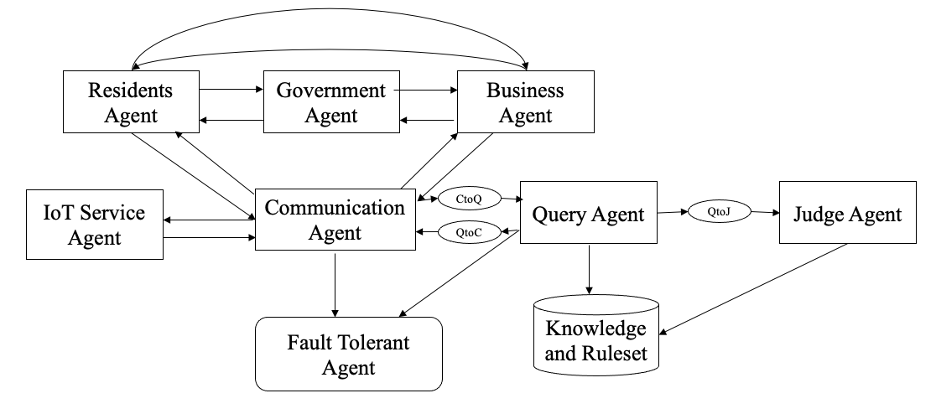}
     \caption{Architecture of ethical judgment multi-agent system}
     \label{fig:architecture_MAS}
 \end{figure}
 
Figure 2. illustrates the architecture for the ethical judge MAS, the Residents, Government and Business agents can interact with each other; the IoT service agent interacts with the three social entity agents through communication agents. The query agent which is for querying about conditions of specific rules can access knowledge and ruleset component and communicate with communication agent. Lastly, the Judge agent makes ethical judgment based on results received from the Query agent, specifically:

\textbf{IoT Service Agent:} This agent gathers data from deployed devices and communicates their operational parameters. It generates instances that reflect device behavior and usage in real time, which are then used to assess compliance with safety and privacy principles.

\textbf{Resident Agent:} This agent represents the interests and rights of residents. It provides information about residents’ preferences, economic status, and concerns regarding IoT applications. This data is crucial for evaluating fairness and privacy issues.

\textbf{Government Agent:} Responsible for overseeing smart city projects, this agent assesses whether the IoT applications comply with regulatory standards and ethical guidelines. It ensures that safety measures are enforced and that the deployment aligns with community welfare.

\textbf{Business Agent:} This agent provides insights into how businesses interact with IoT services. It evaluates whether businesses are leveraging these technologies equitably and in compliance with established ethical standards.

\textbf{Communication Agents:} Facilitate interactions among the Residents, Government, Business, and IoT Service agents, ensuring seamless information exchange.

\textbf{Query Agent:} Responsible for querying specific conditions related to ethical rules. It accesses the knowledge base and ruleset component to retrieve relevant information and communicate findings to the Judge agent.

\textbf{Judge Agent:} The core component that synthesizes the information received from the Query Agent. It evaluates the compliance of IoT applications against established ethical principles and makes decisions regarding any potential violations.
\\ \hspace*{\fill} \\

\subsection{Detailed Steps of How Different Agents Work}

 For the Ethical Judgment Multi-Agent System (MAS) to function effectively in assessing a specific smart city IoT application, each agent must follow a structured process to interact with real-world data. Below is a detailed breakdown of the steps involved for each agent:

\textbf{\underline{IoT Service Agent:}}

•	\textbf{Representing Lists of Parameters:}

-  Identify key parameters relevant to the IoT service, such as device type, location, movement capabilities, and interaction modes.

-  Aggregate this data to form a comprehensive list that reflects the operational characteristics of the IoT application.

•	\textbf{Generating Instances:}

-  Create concrete instances of the IoT service based on the identified parameters. For example, specify a smart traffic light system, its locations, and interaction types with residents and vehicles.
\\ \hspace*{\fill} \\
\textbf{\underline{Residents Agent:}}

•	\textbf{Representing Lists of Parameters:}

-  Collect data on resident demographics, economic status, usage preferences for IoT services, and any authority they hold to address concerns.

•	\textbf{Generating Instances:}

-  Generate instances that represent specific residents or groups of residents, detailing their attributes and interactions with the IoT service.
\\ \hspace*{\fill} \\
\\ \hspace*{\fill} \\
\\ \hspace*{\fill} \\
\\ \hspace*{\fill} \\
\textbf{\underline{Government Agent:}}

•	\textbf{Representing Lists of Parameters:}

-  Define the goals of smart city projects, oversight responsibilities, and safety standards relevant to the IoT service.

•	\textbf{Generating Instances:}

-  Create instances representing government policies or regulations that apply to the deployment and operation of the IoT application.
\\ \hspace*{\fill} \\
\textbf{\underline{Business Agent:}}

•	\textbf{Representing Lists of Parameters:}

-  Identify business types, locations, and their interactions with the IoT service, including how they may benefit from data access.

•	\textbf{Generating Instances:}

-  Produce instances that reflect specific businesses and their parameters in relation to the smart city application.
\\ \hspace*{\fill} \\
\textbf{\underline{Communication Agent}}

•	\textbf{Facilitating Data Communication:}

-  Ensure seamless data transfer between the IoT Service Agent, Residents Agent, Government Agent, and Business Agent.

-  Pass the generated instances of data structures to the Query Agent and Judge Agent for analysis.
\\ \hspace*{\fill} \\
\textbf{\underline{Query Agent}}

•	\textbf{Formalizing Ethical Rules:}

-  Convert ethical principles into a formal ruleset in PVS or assertions in Alloy.

•	\textbf{Applying Theorem Provers/Analyzers:}

-  Use a theorem prover for PVS to check the validity of the rules against the generated instances.

-  Utilize the Alloy Analyzer to evaluate whether the assertions hold true for the created models.

\textbf{\underline{Judge Agent}}

•	\textbf{Receiving Results:}

-  Collect findings from the Query Agent regarding rule compliance and any identified violations.

•	\textbf{Making Final Judgments:}

-  Analyze the results and make informed decisions regarding the ethical compliance of the smart city application.

•	\textbf{Providing Feedback:}

-  Present the judgment outcomes to stakeholders, including recommendations for improvements or necessary actions to address any ethical concerns.
\\ \hspace*{\fill} \\
\\ \hspace*{\fill} \\
\textbf{Overall Workflow} 

1.	Each agent generates and communicates instances of their respective data structures.

2.	The Communication Agent facilitates interaction and data sharing.

3.	The Query Agent assesses compliance by applying ethical rules.

4.	The Judge Agent synthesizes the results and delivers ethical judgments, ensuring the smart city IoT application adheres to established ethical standards.

This structured approach promotes a thorough and systematic evaluation of ethical considerations in smart city applications, fostering accountability and transparency among all stakeholders.
\\ \hspace*{\fill} \\
\subsection{Implementation of Ethical Judgment Framework}

In our exploration of ethical judgment for smart city applications, we attempted to implement models using both Alloy and PVS specification languages. Alloy proved to be particularly effective due to its simplicity and expressiveness, allowing us to model various agents and their interactions efficiently.
\\ \hspace*{\fill} \\
\textbf{Alloy Implementation}
\\ \hspace*{\fill} \\
In Alloy, we constructed models for each social entity: Residents, Government, Business, and IoT Service Agents. Figure 3. shows one example Alloy program which includes definition of smart city entities. Additionally, we formalized our ethical rules as assertions within the Alloy framework. For instance, one critical ethical rule states:

\textit{“The IoT device should not collect and/or analyze user data and make that information public or available to other residents without consent.”}

To test these assertions, we utilize the Alloy Analyzer. By executing the Alloy program, we can quickly determine if any counterexamples exist, indicating a violation of the ethical rules. If the console reports a counterexample, it highlights scenarios where the assertion is invalid, which provides valuable insights into potential ethical breaches.

\begin{figure}[!h]
     \centering
     \includegraphics[width=0.45\textwidth]{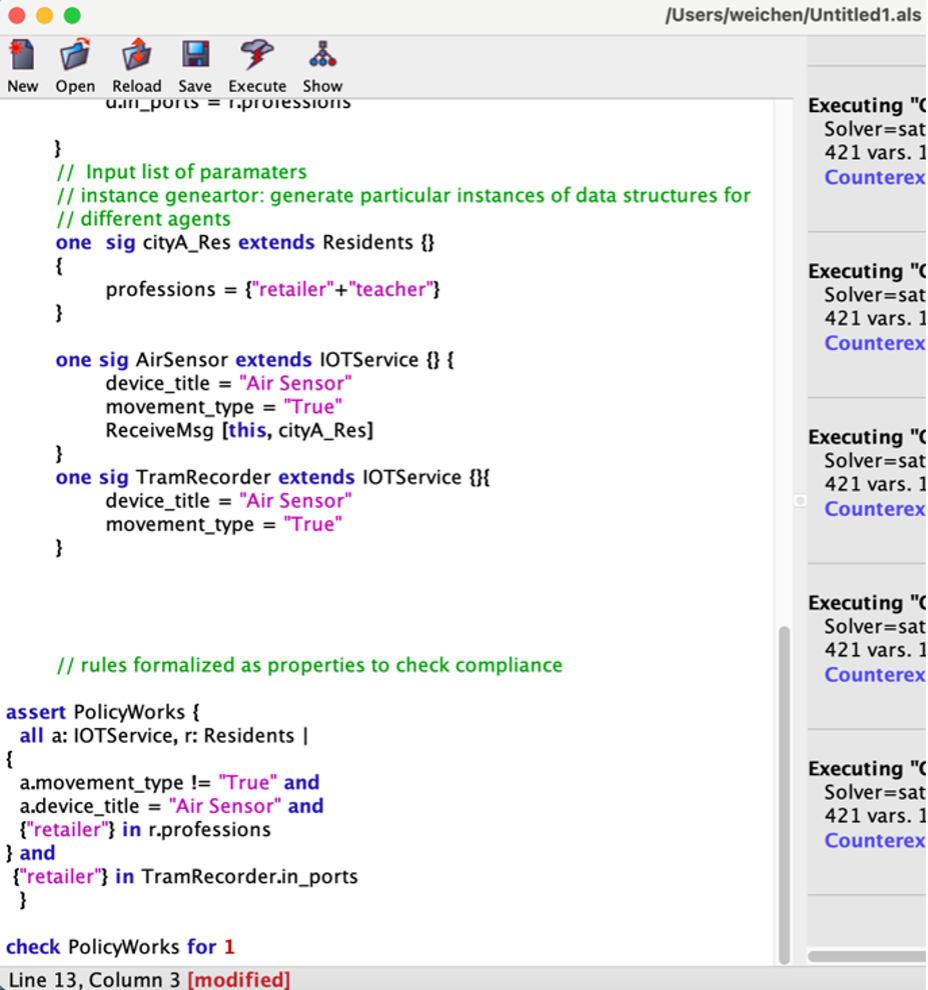}
     \caption{Alloy implementation of agents and rules}
     \label{fig:alloy_implementation}
 \end{figure}

\textbf{Challenges with Data Communication Rules}
\\ \hspace*{\fill} \\
However, we find that certain ethical rules, particularly those involving the communication and transmission of data between social entities, could present challenges in Alloy. These rules often require complex predicates that can become cumbersome to encode directly in Alloy. For example, capturing nuanced consent mechanisms and data-sharing protocols can result in lengthy and intricate predicates, making implementation time-consuming.
\\ \hspace*{\fill} \\
\textbf{Exploring PVS for Enhanced Modeling}
\\ \hspace*{\fill} \\
Given these challenges, we investigated whether PVS could serve as an alternative for encoding these communication-related ethical rules as with its focus on system requirements and more robust logical constructs, PVS offers the potential to express these rules more clearly and concisely. However, we found that PVS is better suited for establishing foundational requirements rather than modeling dynamic data communications.
\\ \hspace*{\fill} \\
\textbf{Integration and Human Judgment}
\\ \hspace*{\fill} \\
Ultimately, while Alloy provides an efficient means for modeling and testing assertions related to ethical compliance, the complexity of certain rules suggests a complementary role for human judgment. In ~\ref{sec:case-study}, we will elaborate on how integrating human oversight can enhance the automatic judgment provided by the Alloy framework, especially in interpreting and enforcing ethical standards that are more complex or context-dependent.

This dual approach of leveraging Alloy for automated analysis while recognizing the importance of human judgment allows for a more comprehensive and nuanced evaluation of ethical considerations in smart city applications.
\\ \hspace*{\fill} \\

\section{Ethical Judgment Case Study and Simulations}
\label{sec:case-study}
This section provides one real-world example of ethical judgment analysis based on smart city applications in the transportation domain. In order to test our framework, for the application we first make comprehensive human-based judgment about all of their potential unethical outcomes to obtain ground truth, then we apply the proposed ethical framework to perform Alloy-based automatic ethical judgment and human judgment to see if the results match and validate if the proposed framework is applicable in real-world context.
\\ \hspace*{\fill} \\
\textbf{Simulation:  FLASH’s Parking Technology}
\\ \hspace*{\fill} \\
\textbf{Description:}~\cite{FLASH} Parkway found a future-ready partner in FLASH, the creator of the first mobility hub operating system. Designed to scale, FLASH’s sophisticated technologies power locations across Parkway portfolio, giving them the tools to secure market leadership and meet new parking and mobility needs in the smart city.

With the installation of 230 devices, from entry/exit kiosks to mini-smart stations and validation kiosks, the new system allowed Parkway facilities to differentiate themselves in a positive way, with an improved seamless customer experience from entry to exit.

For monthly parkers, the process has become completely contactless with Bluetooth and LPR access.

The ability to deploy a variety of contactless integrations (LPR, Bluetooth, pay-to-text, FLASH mobile app) has allowed these facilities to provide a safer experience for all their customers by minimizing touch as much as possible.
\\ \hspace*{\fill} \\
FLASH solutions at Parkways locations include:

•	Parking access and revenue control

•	Real-time analytics and reporting

•	Bluetooth access for monthly parkers

•	Validations

•	Remote mobile app management

•	License Plate Recognition (LPR)

\subsection{Potential Unethical Outcomes (Ground Truth)}

\textbf{Safety Concerns:} Payment processers will impose safety hazard if it requires a user, particularly non-monthly parkers, to hold a payment card and get close to the parking kiosk to pay.

\textbf{Data Misuse: }There is a risk that collected data from real-time analytics and reporting may be shared with third parties for unauthorized use.

\textbf{Privacy Concerns: }Installed camera for recognizing license plates may make users worry about their privacy, and the use of LPR and Bluetooth raises questions about data privacy, as the system collects and processes personal data from users without explicit consent.

\textbf{Inequitable Access: }The parking devices might be primarily installed in neighborhood with more entry/exist traffic flow and not available in other locations, and the service may inadvertently favor users who can afford advanced services, excluding low-income residents.
\\ \hspace*{\fill} \\
\subsection{Data Source}

\textbf{FLASH Parking System Data:} Simulated data on device locations, usage patterns, and properties of technologies to define agent behaviors.

\textbf{Publicly Available Data:} Reports, case studies, and articles for obtaining application context of smart parking.

\textbf{Regulatory \& legal docs:} Relevant legal documents, privacy policies, and regulations related to the system.

\textbf{Mock Data:} Simulated data created due to the difficulty in obtaining real-world operational data or the limited availability of public data.

This study mainly uses mock data to facilitate the study and ensure the	simulation could proceed with limited access to other types of data.

\subsection{Agents Modeling (Simulated MAS)}

Below is a simulated MAS for the FLASH’s parking device at a particular moment during its operation:
\\ \hspace*{\fill} \\
\textbf{\underline{Parking Device Agent}}
\\ \hspace*{\fill} \\
one sig Parking\_device\{

device\_title = “FLASH Parking”,

	deploy\_neighborhood = {“Center City”, ”Fairmount”, ”Queen Village”,  ”Brewerytown”},
 
	movement\_type = “still”,
 
	interaction\_type = “physical”,
 
	risk\_type = “high”,
 
IoT\_usage\_agreement\_violated = “False”,

	\}

\textbf{\underline{Resident Agent}}
\\ \hspace*{\fill} \\
one sig Residents \{

	all\_living\_neighborhoods = {“Center City”, ”Fairmount”, 
 ”Queen Village”, ”Brewerytown”, ”Fishtown” , ”Point Breeze”, ”Old City”, ”North Phily East”, ”South Phily”,  ”Grays Ferry”},
 
	favored\_neighborhoods = {“Center City”, ”Old City”, ”Fishtown”},
 
	economic\_status = {“High”, ”Medium”, ”Low”},
 
	preference\_for\_IoT\_usage = {“Parking Access \& Payment”},
 
	has\_legitimate\_authority = “True”,		
	\}
\\ \hspace*{\fill} \\
\textbf{\underline{Government Agent}}
\\ \hspace*{\fill} \\
one sig Government \{

	gov\_types = “Transportation Dept”
 
	IoT\_project\_goals : {“Improve City Infrastructure ”, ”Technological Reform”},

	oversight\_IoT\_safety = “True”,
 
	enforce\_safety\_standards = “True”,		
 
	\}
\\ \hspace*{\fill} \\
\textbf{\underline{Business Agent}}
\\ \hspace*{\fill} \\
one sig BA \{

	neighborhoods = {“Center City”, ”Fairmount”, ”Queen Village”,  ”Brewerytown”, ”Old City”+”Fishtown”},
 
	type\_of\_business = {“Cafe”,  ”Gas Station” }	
 
	\}

\subsection{Automatic Ethical Judgment}

After implementing the ethical judgment framework using Alloy, we found counterexamples indicating potential violations of ethical principles:

\textbf{Principle P1 (Safety): }The ethical dimension concerning safety may be compromised if the parking kiosk requires users to physically interact with the device for payment. This action involves physical movement of the residents entity in a car and imposes a risk of transmission of harmful substances or information due to contact.

\subsection{Human Ethical Judgment}

In addition to the automatic judgment, human assessment was conducted for the following principles:
\\ \hspace*{\fill} \\
\textbf{Principle P4, P5, P6 (Privacy):} Investigated if user data collected from the parking kiosk or mobile apps could be made available to government agencies, other residents, or third-party businesses without proper consent.
\\ \hspace*{\fill} \\
\textbf{Principle P7 (Fairness):} Examined if residents were treated equitably by businesses by interacting with them via the parking device.
\\ \hspace*{\fill} \\
Our findings are:
\\ \hspace*{\fill} \\
\textbf{Principles P4 and P6:}
Violations were identified under conditions where user data collected through the parking system was shared with government agencies or third parties without obtaining explicit consent from the users. This raises significant ethical concerns regarding user privacy and data protection.
\\ \hspace*{\fill} \\

\subsection{Validate Judgment Results with Ground Truth}
 
We found the ethical concerns identified through a combination of both methods, such as physical interaction with payment kiosks (Safety), unauthorized data sharing (Privacy), and data collection without consent (Privacy), align with the ground truth concerns of the parking system case, indicating that the proposed framework can be applied to effectively detect ethical violations in smart city applications.

\section{Discussions on the Judgment Framework}
\label{sec:discussion}
In this section, we explore key aspects of the proposed framework, drawing insights from the right-based ethical rules and the ethical judgment examples provided earlier.

\subsection{Generalizability and Addition of New Rules}

A critical goal of our framework is to ensure that the predefined ethical principles and rules are comprehensive and applicable across various smart city applications. To assert that our set of 13 ethical principles is sufficiently complete, we need to demonstrate that the rules we formulated can effectively prevent every conceivable unethical outcome related to each right in the rights-based framework. However, given the complexity and variability of potential unethical situations, we acknowledge that while our ethical rules can identify general violations, they may not capture every specific scenario. This raises concerns about the completeness and generalizability of our ruleset.

To enhance the framework, new rules can be introduced by referring to the Alloy specifications of agents and utilizing the defined properties to articulate additional rules. This adaptability is essential for keeping pace with evolving ethical considerations in smart city applications.

\subsection{Automatic Judgment vs. Human Judgment}

The integration of Alloy for automatic judgment presents several advantages over human judgment, especially concerning efficiency and reliability:

\textbf{Efficiency:} Automatic judgment systems can execute rule checking almost instantaneously once the necessary parameters are input, making them a time-effective solution for ethical assessments.

\textbf{Reliability:} Logical rules can encapsulate complex relationships among multiple entities, which might be challenging for human judges to evaluate consistently. The precise nature of logical expressions ensures that rules can be systematically applied without subjective bias.

\textbf{Reusability:} As part of a knowledge base, logical rules can be modified and reused, allowing for continual improvement of the ethical judgment framework as new scenarios and insights emerge.

Despite these advantages, there remain areas where human judgment is indispensable, particularly in interpreting nuances of ethical implications that may not be fully captured by automated systems.

\subsection{Application of the Framework in Practice}
The practical application of the ethical judgment framework serves two primary purposes:

\textbf{For Smart City Developers:} The framework provides a mechanism for developers to assess their applications for potential unethical outcomes. This proactive approach enables developers to identify and mitigate risks before deployment.

\textbf{For Government Regulators: }The framework can also assist regulators in evaluating whether smart city applications comply with established ethical principles. However, it is crucial that developers furnish regulators with detailed specifications to facilitate effective assessments.

In practice, a user-friendly interface would significantly enhance usability, allowing judges to select specific rules for evaluation and execute them either independently or from a broader perspective.

\subsection{Social Control and Technical Control}
An effective multi-agent system (MAS) for smart cities should embody both social control over ethical principles and technical control over its agents. Our framework supports social control by allowing modifications or additions to ethical rules in response to shifting ethical practices within the smart city context.

Simultaneously, the technical control aspect is addressed through the MAS's ability to identify which agents and their properties are implicated in detected violations. This identification allows for targeted actions, whether through redesign, adjustment of functionalities, or enhanced oversight.
\\ \hspace*{\fill} \\
\section{Conclusion and Future Work}
This paper provides a comprehensive overview of the approaches for analyzing and judging ethical issues that may raise by IoT technologies in smart cities. The described framework adopts multi-agent approach and includes both an ethical rules table designed by human and a multi-agent judgment framework. By applying the framework to real-world examples of smart city applications, the validity of the framework can be examined, and more fine-grained ethical principles can be revealed to guide ethical smart city development.

The proposed ethical judgment framework offers a robust foundation for evaluating ethical implications in smart city applications. By balancing automatic and human judgment, supporting adaptability through new rule integration, and fostering both social and technical control, the framework positions itself as a vital tool for ensuring ethical compliance in the rapidly evolving landscape of smart cities. Continued evaluation and refinement of the framework will be essential to address emerging challenges and uphold the rights and well-being of residents. Additionally, the proposed framework could be integrated into design or deployment tools for emerging smart city applications.
\\ \hspace*{\fill} \\

\bibliographystyle{IEEEtran}
\bibliography{references}

\end{document}